\documentclass{scspaperproc}

\usepackage{latexsym}
\usepackage{graphicx}
\usepackage{mathptmx}

\usepackage{amsmath}
\usepackage{amsfonts}
\usepackage{amssymb}
\usepackage{amsbsy}
\usepackage{amsthm}
\usepackage[pdftex,colorlinks=true,urlcolor=blue,citecolor=black,anchorcolor=black,linkcolor=black]{hyperref}

\usepackage{hyphenat}
\newtheoremstyle{scsthe}% hnamei
{8pt}% hSpace abovei
{8pt}% hSpace belowi
{\it}% hBody fonti
{}% hIndent amounti1
{\bf}% hTheorem head fontbf
{.}% hPunctuation after theorem headi
{.5em}% hSpace after theorem headi2
{}% hTheorem head spec (can be left empty, meaning `normal')i

\theoremstyle{scsthe}

\begin{document}

%***************************************************************************
% AUTHOR: AUTHOR NAMES GO HERE
% FORMAT AUTHORS NAMES Like: Author1, Author2 and Author3 (last names)
%
%		You need to change the author listing below!
%               Please list ALL authors using last name only, separate by a comma except
%               for the last author, separate with "and"
%

% setting up general page style
\pagestyle{fancyplain}

% setting up page style of first page
\thispagestyle{plain}
\firstPageHead{}

% setting up running header (authors) of subsequent pages
\chead{\fancyplain{}{\itshape Dahal, Ioup, Arifuzzaman, and Abdelguerfi \vspace{8pt}}}

% setting up seperation parameters
%\headsep=72pt
\rhead{}
\cfoot{}
\renewcommand{\headrulewidth}{0pt} % (renewcommand needed in fancyhdr to remove top decorative line)
%\headrulewidth=0pt  % ("setlength" needed in fancyheading to remove top decorative line)

%%%%%%%%%%%%%%%%%%%%%%%%%%%%%%%%%%%%%%%%%%%%%%%%%%%%%%%%%%%%%%%%%%%%%%%%%%%%%%
%                                                                            %
%     THESE COMMANDS ARE REQUIRED TO WORK WITH SCSPROC.BST TO MAKE BIBLIO    %
%                                                                            %
%%%%%%%%%%%%%%%%%%%%%%%%%%%%%%%%%%%%%%%%%%%%%%%%%%%%%%%%%%%%%%%%%%%%%%%%%%%%%%
\makeatletter
\let\@internalcite\cite
\def\cite{\def\@citeseppen{-1000}%
    \def\@cite##1##2{(##1\if@tempswa , ##2\fi)}%
    \def\citeauthoryear##1##2##3{##1 ##3}\@internalcite}
\def\citeNP{\def\@citeseppen{-1000}%
    \def\@cite##1##2{##1\if@tempswa , ##2\fi}%
    \def\citeauthoryear##1##2##3{##1 ##3}\@internalcite}
\def\citeN{\def\@citeseppen{-1000}%
%  Pierre L'Ecuyer's fix for multiple cite bug
%  Added by Paul J Sanchez on 4 October 2001
%   \def\@cite##1##2{##1\if@tempswa , ##2)\else{)}\fi}%
%   \def\citeauthoryear##1##2##3{##1 (##3}\@citedata}
    \def\@cite##1##2{##1\if@tempswa, ##2)\else{}\fi}%
    \def\citeauthoryear##1##2##3{##1 (##3)}\@citedata}
\def\citeA{\def\@citeseppen{-1000}%
    \def\@cite##1##2{(##1\if@tempswa , ##2\fi)}%
    \def\citeauthoryear##1##2##3{##1}\@internalcite}
\def\citeANP{\def\@citeseppen{-1000}%
    \def\@cite##1##2{##1\if@tempswa , ##2\fi}%
    \def\citeauthoryear##1##2##3{##1}\@internalcite}
\def\shortcite{\def\@citeseppen{-1000}%
    \def\@cite##1##2{(##1\if@tempswa , ##2\fi)}%
    \def\citeauthoryear##1##2##3{##2 ##3}\@internalcite}
\def\shortciteNP{\def\@citeseppen{-1000}%
    \def\@cite##1##2{##1\if@tempswa , ##2\fi}%
    \def\citeauthoryear##1##2##3{##2 ##3}\@internalcite}
\def\shortciteN{\def\@citeseppen{-1000}%
%  Pierre L'Ecuyer's fix for multiple cite bug
%  Added by Paul J Sanchez on 2 September 2002
%  should have caught this last year...
%   \def\@cite##1##2{##1\if@tempswa , ##2)\else{)}\fi}%
%   \def\citeauthoryear##1##2##3{##2 (##3}\@citedata}
% Shane G. Henderson fix for extra right bracket at end of optional material June 8, 2005
%    \def\@cite##1##2{##1\if@tempswa, ##2)\else{}\fi}%
    \def\@cite##1##2{##1\if@tempswa, ##2\else{}\fi}%
    \def\citeauthoryear##1##2##3{##2 (##3)}\@citedata}
\def\shortciteA{\def\@citeseppen{-1000}%
    \def\@cite##1##2{(##1\if@tempswa , ##2\fi)}%
    \def\citeauthoryear##1##2##3{##2}\@internalcite}
\def\shortciteANP{\def\@citeseppen{-1000}%
    \def\@cite##1##2{##1\if@tempswa , ##2\fi}%
    \def\citeauthoryear##1##2##3{##2}\@internalcite}
\def\citeyear{\def\@citeseppen{-1000}%
    \def\@cite##1##2{(##1\if@tempswa , ##2\fi)}%
    \def\citeauthoryear##1##2##3{##3}\@citedata}
\def\citeyearNP{\def\@citeseppen{-1000}%
    \def\@cite##1##2{##1\if@tempswa , ##2\fi}%
    \def\citeauthoryear##1##2##3{##3}\@citedata}
%
% \@citedata and \@citedatax:
%
% Place commas in-between citations in the same \citeyear, \citeyearNP,
% \citeN, or \shortciteN command.
% Use something like \citeN{ref1,ref2,ref3} and \citeN{ref4} for a list.
%
\def\@citedata{%
    \@ifnextchar [{\@tempswatrue\@citedatax}%
                  {\@tempswafalse\@citedatax[]}%
}

\def\@citedatax[#1]#2{%
\if@filesw\immediate\write\@auxout{\string\citation{#2}}\fi%
  \def\@citea{}\@cite{\@for\@citeb:=#2\do%
    {\@citea\def\@citea{, }\@ifundefined% by Young
       {b@\@citeb}{{\bf ?}%
       \@warning{Citation `\@citeb' on page \thepage \space undefined}}%
{\csname b@\@citeb\endcsname}}}{#1}}%

% don't box citations, separate with ; and a space
% also, make the penalty between citations negative: a good place to break.
%
\def\@citex[#1]#2{%
\if@filesw\immediate\write\@auxout{\string\citation{#2}}\fi%
  \def\@citea{}\@cite{\@for\@citeb:=#2\do%
    {\@citea\def\@citea{, }\@ifundefined% by Young
       {b@\@citeb}{{\bf ?}%
       \@warning{Citation `\@citeb' on page \thepage \space undefined}}%
{\csname b@\@citeb\endcsname}}}{#1}}%

% (from apalike.sty)
% No labels in the bibliography.
%
\def\@biblabel#1{}
\makeatother

\newdimen\bibindent
\bibindent=.25in

% SEC: was \def\thebibliography#1{\section*{\refname\@mkboth
% SEC: was   {\uppercase{\refname}}{\uppercase{\refname}}}\list
\def\thebibliography#1{\section*{\refname}\list
   {}{\settowidth\labelwidth{[#1]}
   \leftmargin \bibindent
   \itemindent -\bibindent
   \listparindent \itemindent
	 \itemsep 4pt
   \parsep 0pt
   \usecounter{enumi}}
   \def\newblock{}
   \sloppy
   \sfcode`\.=1000\relax}
           % Set up BiBTeX macros

% needed to make the tex document look more like the word counterpart :-(
\setlength{\baselineskip}{12.7pt}

% AUTHOR: Uncomment ONE of these correct conference names.
\def\SCSconferenceacro{SpringSim}
\title{Distributed Streaming Analytics on Large-scale Oceanographic Data using Apache Spark}

% AUTHOR: Enter the authors of the article, see end of the example document for further examples
\author{
Janak Dahal \\ [12pt]
Dept. of Computer Science, \\ 
University of New Orleans,	\\ New Orleans, Louisiana 70148  USA\\
jdahal@uno.edu\\
% Multiple authors are entered as follows.
% If they share the same affiliation, they must be allocated within the same block
% Every author name must be separated by \\ but the last one within the block, that will be separated by \\ [12pt]
% You may also need to adjust the titlevbox size in the preamble - search for titlevboxsize
\and
Elias Ioup \\ [12pt]
US Naval Research Laboratory, \\ Stennis Space Center, MS 39529 USA \\
elias.ioup@nrlssc.navy.mil\\
\and 
Shaikh Arifuzzaman \\ [12pt]
Dept. of Computer Science, \\ 
University of New Orleans,	\\ New Orleans, Louisiana 70148  USA\\
smarifuz@uno.edu\\\\
\and
Mahdi Abdelguerfi \\ [12pt]
Dept. of Computer Science, \\ 
University of New Orleans,	\\ New Orleans, Louisiana 70148  USA\\
mahdi@cs.uno.edu\\
}

\maketitle

\section*{Abstract}

Real-world data from diverse domains require real-time scalable analysis. Large-scale data processing frameworks or engines such as Hadoop fall short when results are needed on-the-fly. Apache Spark's streaming library is increasingly becoming a popular choice as it can stream and analyze a significant amount of data. In this paper, we analyze large-scale geo-temporal data collected from the USGODAE (United States Global Ocean Data Assimilation Experiment) data catalog, and showcase and assess the ability of Spark stream processing. We measure the latency of streaming and monitor scalability by adding and removing nodes in the middle of a streaming job. We also verify the fault tolerance by stopping nodes in the middle of a job and making sure that the job is rescheduled and completed on other nodes. We design a full-stack application that automates data collection, data processing and visualizing the results. We also use Google Maps API to visualize results by color coding the world map with values from various analytics.

\textbf{Keywords:} Streaming analytics, Apache Spark, Real-time processing, Hadoop, Temporal data, Scalable methods.
%% AUTHOR:
% This is a list of no more than five keywords that will identify your paper in indices and databases (required).
% Do not use the words “computer”, “simulation”, “model”, or “modeling”, since these are all assumed.

\section{ INTRODUCTION}

\noindent Processing and analyzing data in real time can be a challenge because of its size. In the current age of technology, data is produced and continuously recorded by a wide range of sources. According to a marketing paper published by IBM in 2017, as of 2012, 2.5 quintillion bytes of data was generated every day, and 90\% of the world's data was created since 2010 \cite{RN1}. With new satellites, sensors, and websites coming into existence every day, data is only bound to grow exponentially. The number of users interacting with theses mediums are producing data at an enormous rate \cite{triangle_dynamic}. With the Internet reaching to new nooks and corners of the world, sources of potential data are ever-growing. As more data keep coming into existence, the necessity of a system that can analyze it in real-time becomes even more imminent. Although the concept of batch processing (using multiple commodity machines in a truly distributed setting) was a revolution when it first came into existence, it might not be a complete solution to the need for real-time processing. Such on-the-fly processing has applictions in many areas such as banking, marketing, and social media. For example, identifying and blocking fraudulent banking transactions require quick actions by processing vast amounts of data and producing quick results. Sensitive and illegal posts on social media can be quickly removed to nullify the adverse effects on its users. Weather data, like the one used in this research, can be analyzed in real time to detect or predict different climatic conditions.

\section{ BACKGROUND}

The notion of using commodity machines as a computational power came into existence with the advent of Google File System (GFS). It introduced a distributed file system that excelled in performance, scalability, reliability, and availability \cite{RN2}. As this truly distributed and replicated file system became rigidly stable, the next step in the ladder was to be able to process the data stored in it. For this, Google introduced MapReduce as a programming model \cite{RN3}. This new parallel programming model demonstrated the ability to write small programs (map and reduce classes) for processing big data. It introduced the concept of offloading computation to the data itself and thus nullifying the effect of network bottleneck on batch processing by not having to move the input data between nodes. Hadoop is the most popular MapReduce framework today, but it has its limitations. The most prominent shortcoming of Hadoop lies in the iterative data-processing \cite{RN4}. To extend Hadoop beyond conventional batch processing requires various third-party libraries. Storm can be used along with Hadoop to accomplish real-time processing \cite{RN5}. Other libraries such as Hive, Giraph, HBase, Flume, and Scalding are designed to tackle specific operations, e.g., querying and graphing. Managing these different libraries can be time-consuming from a development point of view.

With Hadoop's limitations in mind, a new framework called Spark was designed that would reuse a working set of data across multiple operations \cite{RN4}. The more iterative a computation is, the more efficient is the job running on Apache Spark. Spark streaming library has become widely popular to run real-time processing jobs. This library allows applications to stream data from different sources \cite{RN6}. Some of the most popular streaming sources include Kafka, Flume, Twitter, and HDFS. Data can be streamed into the streaming job from one or more sources and unified into a single stream. For the application designed for this paper, data is streamed from the Hadoop File System (HDFS).

%Further, the processing libraries are directly written over its core that provide streaming, querying and graphing operations [4]. 
\section{ APACHE SPARK}

Introduced in a paper published in 2010, Spark is a cluster computing framework that uses a read only collection of objects called Resilient Distributed Datasets (RDDs) that let users perform in-memory calculations on large clusters \cite{RN7}. RDDs are fault-tolerant, parallel data structures which makes it possible to explicitly persist intermediate results in memory, control their partitioning to optimize data placement, and manipulate them using a rich set of operators \cite{RN7}. As the intermediate results are stored in memory, iterative analytics such as PageRank calculation, k-means clustering, and linear regression become much more efficient in Spark compared to Hadoop \cite{RN5}. %\textbf{\underbar{}}

\subsection{ Resilient Distributed Data (RDD)}

\noindent RDD is defined as a collection of elements partitioned across different nodes in a cluster that can be operated on in parallel \cite{RN7}. From a user's point of view, it looks like a data structure, but behind the scenes, it performs all the operations necessary to run in a distributed framework. Failures across large clusters are inevitable; thus, the RDDs in Spark were designed with fault tolerance in mind. Since most of the operations in Spark are lazy (no operations are run on data unless an action, e.g., collect, reduce, etc., is called), the operations on RDDs are stored in the form of a Directed Acyclic Graph (DAG). A DAG is a collection of functional lineage such as map and filter. Such awareness of the functional lineage makes it possible for Spark to handle node failures gracefully \cite{RN7}.  These RDDs drive the streaming framework in Apache Spark. They have the following properties that make sure the Apache Spark Streaming maintains its integrity:

\textbf{Replicated.} RDDs are split between various data nodes in a cluster. Replicas are also spread across the cluster to make sure that the system can recover from any aftermath of the node crash. Processing occurs on nodes in parallel, and all RDDs are stored in memory on each node.

\textbf{Immutable.} When an operation is performed on an RDD, the original RDD is not changed. Instead, a new RDD is created out of that operation \cite{RN7}. Only two operations are performed on an RDD namely \textit{transformation} and \textit{action}. A \textit{transformation} transforms the RDD into a new one whereas an \textit{action} gets data from the RDD.

\textbf{Resilient.} Resiliency pertains to the replication of data and storing the lineage of operation on RDDs. When a worker node crashes, the state of the RDD can be regenerated by running the same set of transformations to reach the current state of the RDD \cite{RN7}.

\subsection{ Apache Spark Streaming}

\begin{figure}[!ht]
	\begin{center}
		\includegraphics[scale=0.95]{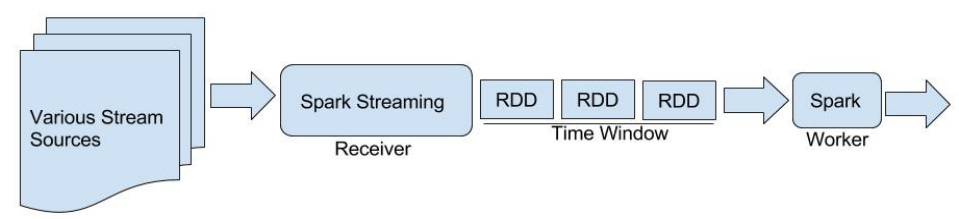}
		\caption{ \small Outline of Apache streaming framework used in this paper.}
	%	\vspace{-0.2in}
		\label{fig:apachestream}
	\end{center}
\end{figure}

In many real-world applications, time-sensitive data can often get stale very quickly. Thus, to make the most of such data, it must be analyzed on time. For example, if a banking website starts generating piles of 500 errors, the potential of an incoming request crashing the server must be evaluated in real time. Traditional MapReduce is not a viable solution for such cases as it is mostly suited for offline batch processing where results are not associated with any latency \cite{RN4}.  If the input data is repeatedly produced in discrete sets, multiple passes of the map and reduce tasks would create overhead which can be eliminated by using Spark instead. Apache Spark Streaming lets the program store results in an intermediate format in memory, and when new data arrives as another discrete set, it is batched to perform transformations on them quickly and efficiently \cite{RN4}. Figure \ref{fig:apachestream} outlines the Apache streaming framework.  

%%\noindent \includegraphics*[width=3.84in, height=0.88in, keepaspectratio=false]{image1}

%\noindent \textit{Figure 1:Outline of Apache Streaming}

Data can be streamed into Apache Spark streaming framework from various sources like Kafka, Flume, Twitter, and HDFS \cite{RN8}. A receiver must be instantiated and hooked up with the streaming source to start the flow of data. One receiver can only stream data from one input source, and if we have multiple stream sources, then we can union them so that they can be processed as a single stream \cite{RN9}. Once the receiver starts receiving the data from the streaming source, Spark stores the data into a series of RDDs delineated by a specified time window. After this time, the data is passed into the spark core for processing. To start any Spark streaming job, there needs to be at least two cores, one that receives the data as stream and one that processes the data.

\section{ Streaming Analytics on Large-scale Ocean Data}

%The goal of this application is to be able to run queries on a large dataset and produce results in a certain amount of time which is a magnitude of times faster than running the application in a traditional batch processing fixture. 
We develop an application to run queries on a large oceanographic dataset and produce results on the fly. Apache Spark is chosen for a platform to write the application because of its streaming library. We stream data into the streaming job from HDFS. We collected data from United States Global Ocean Data Assimilation Experiment (USGODAE) data catalog and then processed and stored in the HDFS. The application streams new data within the configurable window of time and run transformations and actions to generate results. %Various steps were involved in the process to accomplish such an application.

\subsection{Setup and Configuration}

%\textbf{ Setting up HDFS.} 
Although Hadoop is not required to run Spark, we installed it because our application reads data from HDFS. Hadoop was first installed on a single node setting, and then other nodes were added one at a time. Each time a node was added, the sample MapReduce tasks were run to make sure that the job was making use of all the nodes. Five nodes with identical computational power were used to create the cluster. %Fig. \ref{fig:hadoopconfig} summarizes the final Hadoop configuration.  

%Configuration of each node is presented in Fig. \ref{fig:nodeconfig}. Further, 

\iffalse
\begin{figure}[!ht]
	\begin{center}
		\includegraphics[scale=0.75]{Figs/Fig2_nodeconfig.pdf}
		\caption{ \small Node CPU configuration.}
		\vspace{-0.2in}
		\label{fig:nodeconfig}
	\end{center}
\end{figure}
\fi
%%\noindent \includegraphics*[width=2.24in, height=2.57in, keepaspectratio=false]{image2}

%\noindent \textit{Figure 2:Node CPU configuration} 

%\noindent 

%%\includegraphics*[width=5.02in, height=3.32in, keepaspectratio=false]{image3}
\iffalse
\begin{figure}[!ht]
	\begin{center}
		\includegraphics[scale=0.75]{Figs/Fig3_hadoopconfig.pdf}
		\caption{ \small Hadoop Cluster Configuration.}
		\vspace{-0.2in}
		\label{fig:hadoopconfig}
	\end{center}
\end{figure}
\fi

%\noindent \textit{Figure 3:Hadoop Cluster Configuration}

\begin{figure}[!ht]
	\begin{center}
		\includegraphics[scale=0.95]{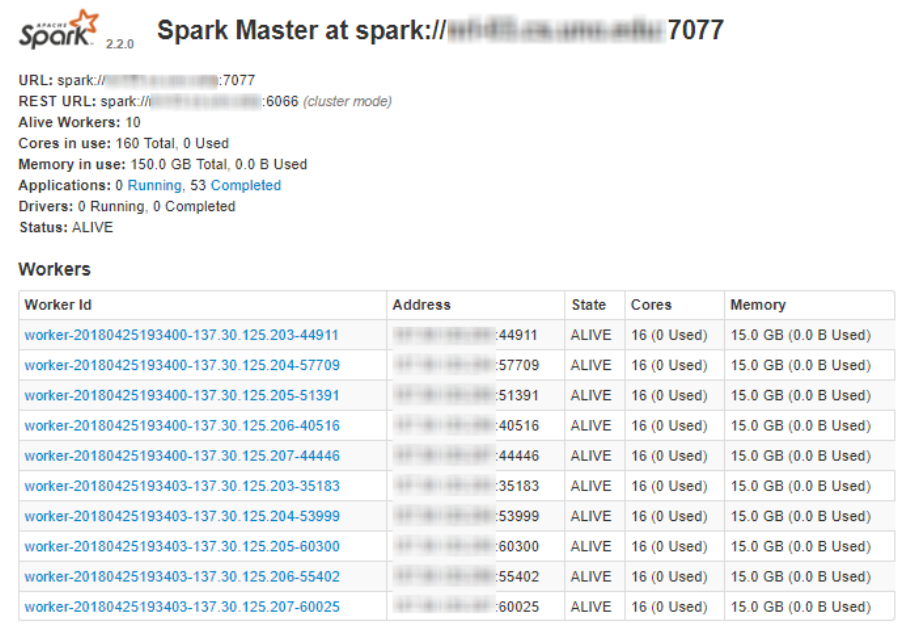}
		\caption{ \small Apache Spark system configuration used for our work. The IP addresses are hidden for privacy reason.}
		%\vspace{-0.2in}
		\label{fig:sparkconfig}
	\end{center}
\end{figure}

We install Apache Spark along with SBT and Scala. SBT is used to build the Scala projects. Scala is used as the programming language of choice to write streaming jobs. We install Spark in the same way as Hadoop by starting with a single node and adding one node at a time. Two workers instances (SPARK\_WORKER\_INSTANCES=2) ran on each terminal to utilize dual CPUs. Each worker is set up to utilize up to 15GB memory (SPARK\_WORKER\_MEMORY=15GB) and up to 16 cores (SPARK\_WORKER\_CORES=16). We set up Hadoop File System(HDFS) on each of the nodes.  YARN, a resource manager and a dashboard to visualize and summarize the metrics, runs on the driver node. We set up REPL environment or Spark-shell in each node to make sure that the debugging is swift when a transformation needs to be performed on a set of data. Figure \ref{fig:sparkconfig} summarizes the Apache Spark installation.

%%\noindent \includegraphics*[width=3.62in, height=2.49in, keepaspectratio=false]{image4}

%\noindent \textit{Figure 4:Apache Spark Configuration}

\subsection{Description of Datasets}

%\paragraph{ Data Files}

Data is generated every 6 hours by an oceanographic model (NAVGEM-Navy Global Environmental Model) that predicts various environmental variables for the next 24 hours to 180 hours. The number of output files from the model depends on the type of variable. The data is generated for $198$ different variables which cover the entire world with a precision interval of $0.5$ degrees. The model generates multiple files with the results, and each file contains only data for a single variable. The complete set of data for ten years is about $110$ TB, but we have only about $4.5$ TB disk space available in the distributed file storage. Therefore, we include only four variables for our experiments: ground sea temperature, pressure, air temperature, and wind speed. We use Panoply [10] as a GUI to visualize the input data and resulting data.

\iffalse
\textbf{Data Format.} The file name for a model result is in the following format: ProductKey$\mathrm{{}^\circ}$AAAA\_BBBB\_CCCCCDDEEFFFFFFFFFF\_ GGGG\_HHHHHH- IIIIIIJ...J

\begin{figure}[!ht]
	\begin{center}
		\includegraphics[scale=0.75]{Figs/Fig5_datafile.pdf}
		\caption{ \small Data file name specifications.}
		\vspace{-0.2in}
		\label{fig:datafile}
	\end{center}
\end{figure}

\fi
%%\noindent \includegraphics*[width=3.83in, height=2.69in, keepaspectratio=false]{image5}

%\noindent \textit{Figure 5:Data file name specifications}

Our datasets cover the entire world, so the size of the data array is 361x720, where 361 represents all latitude points from -90 degrees north to +90 degrees north with $0.5$ degree increments, whereas 720 represents all longitude points from 360 degrees east to 0 degrees NE with $0.5$ degree increments as well.

\textbf{Procedure for Data Collection.} We collect our data using the following steps.

\begin{enumerate}
	\item  A Java program would download the data into the filesystem. The parameter type (Part J) in filename above was used to choose the files before downloading them.
	
	\item  NCAR Command Language was used to convert the data from GRIB (General Regularly-distributed Information in Binary form) format \cite{RN11} into the NetCDF3 data format. % (This changed the variable name by abbreviating them, so an enum class was written to map the abbreviations with the original variable names).
	
	\item  CDO (Climate Data Operators \cite{RN12}, written by the Max Planck Institute for Meteorology) was used to merge the data files so each file could contain data for multiple variables.
	
	\item  Files were copied to the HDFS using standard HDFS commands.
\end{enumerate}
We wrote a bash script to automate the above steps and make them seamless. 

\subsection{SciSpark}

Our application extends the functionality of the SciSpark \cite{RN13} project by changing its open source code as needed. SciSpark library facilitates the process by mitigating the need to write wrapper classes to represent GRIB. The library provides a class called SciTensor that represented NetCDF data and implemented all basic mathematical operations such as addition, subtraction, and multiplication. We add new functions to SciTensor class to calculate maximum (\textit{max}), minimum (\textit{min}), and standard deviation. Other significant changes included logic to account for missing variables in a dataset. For multivariable analysis, we added relevant logic to create unique names for \textit{x} and \textit{y} axes when creating NetCDF result file with more than one variable. We create RDDs using SciTensor library and feed into the spark streaming queue.

\subsection{ Application in Use}

\begin{figure}[!ht]
	\begin{center}
		\includegraphics[scale=0.85]{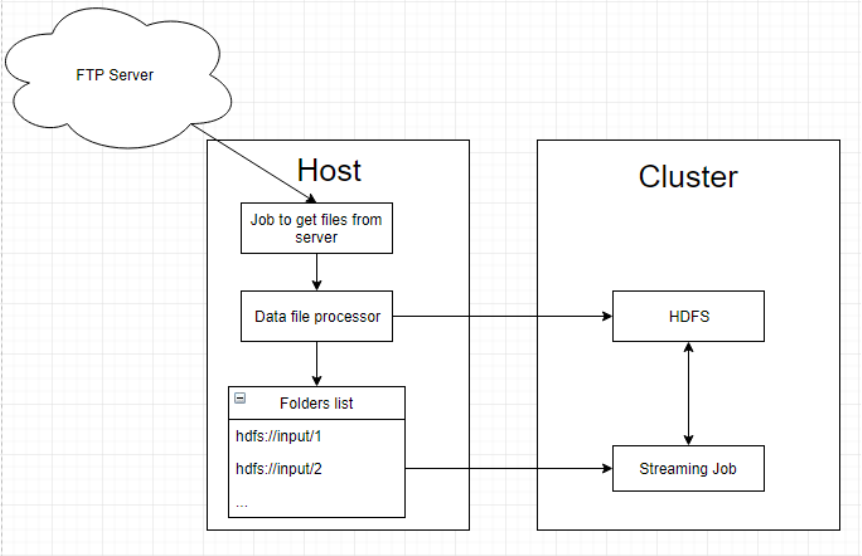}
		\caption{A simplistic overview of our entire application. We design a full-stack application that automates data collection, data processing and visualizing the results. }
		\label{fig:appoverview}
	\end{center}
\end{figure}

Our application streams new files from a location in HDFS and writes the results back to HDFS. The job runs with a configurable time window and performs transformations and actions on all the RDDs accumulated during that time-frame. We use QueueStream API in Apache Spark to read the stream of new RDDs inside the streaming job. New RDDs are represented as a Discretized Stream (DStream) of type SciTensor. Spark Streaming API defines DStream as the fundamental abstraction in Spark Streaming and is a continuous sequence of RDDs (of the same type) \cite{RN14}. Figure \ref{fig:appoverview} summarizes the outline of our application.

%%\noindent \includegraphics*[width=3.42in, height=2.21in, keepaspectratio=false]{image6}

%\noindent \textit{Figure 6:Application Overview}

A scheduled job running on the host runs every hour to download new data from the FTP server. After the download is completed, the data is processed and uploaded to HDFS. The streaming job running on the cluster processes these new files and update the result. The website running on a separate server polls the result file and visualizes the data using Google Maps.

\section{Performance Evaluation}
Statistical analysis is one of the standard operations that programmers use in Apache Spark Streaming to generate summary results in real-time. We analyze certain properties of a streaming framework below. We also evaluate scalability and facult tolerance.

\subsection{ Complexity of the operation}

We evaluate and measure two significant steps in the streaming process namely \textit{transformation} and \textit{action}. We design multiple mathematical queries of varying complexity and run jobs to measure the performance of Apache Spark Streaming. For example, \textit{average}, \textit{maximum} and \textit{minimum} are more straightforward mathematical operations, whereas \textit{standard deviation} can be regarded as a more complex one. We perform the following statistical analyses: mean, max, min, and standard deviation. Once a user submits the streaming job, it cannot be changed for the lifetime of that job. The input sizes per streaming window for each job were approximately 180MB, 500MB, 1GB, and 2GB. The streaming window was set as 6 hours for the streaming process because the input data is produced by the model every 6 hours.

\textbf{Variation of Each Statistical Analysis.} Since GRIB1 data represents values in $361x720$ 2D arrays and the values are scattered across multiple files, to calculate an aggregate for each index, same indices across multiple files were aggregated. To calculate aggregate results for each latitude and longitude points, 361 and 720 more values in each file needed to be aggregated respectively. Moreover, calculating one single aggregate result for all the values across all the files increased the operation complexity as it had to aggregate more values. The variation in statistical analysis in ascending order of complexity is listed as follows: (i) one result for each combination of latitude and longitude points, (ii) one result for each latitude point, (iii) one result for each longitude point, and (iv) one single aggregate result for all data points.

Table \ref{table:statan1} shows the average execution time for each variation of all four statistical analyses. It shows that the complexity of operation is directly proportional to the execution time. More transformations were required on data when running with variation 2, 3 and 4. Each additional transformation increased the length of the DAG and thus increased the execution time.

\begin{table}
	\caption{Result for statistical analysis}	
	\begin{center}
		\centering
	\begin{tabular}{|p{0.6in}|p{0.6in}|p{0.6in}|p{0.6in}|} \hline 
		Variation & Dataset Size & DAG Length & Execution time (s) \\ \hline 
		1 & 180MB & 5 & 22  \\ \hline 
		& 500MB & 5 & 37 \\ \hline 
		& 1GB & 5 & 49 \\ \hline 
		& 2GB & 5 & 81 \\ \hline 
		2 & 180MB & 6 & 23 \\ \hline 
		& 500MB & 6 & 41 \\ \hline 
		& 1GB & 6 & 51 \\ \hline 
		& 2GB & 6 & 101 \\ \hline 
		3 & 180MB & 6 & 22 \\ \hline 
		& 500MB & 6 & 43 \\ \hline 
		& 1GB & 6 & 60 \\ \hline 
		& 2GB & 6 & 117 \\ \hline 
		4 & 180MB & 7 & 42 \\ \hline 
		& 500MB & 7 & 87 \\ \hline 
		& 1GB & 7 & 133 \\ \hline 
		& 2GB & 7 & 278 \\ \hline 
	\end{tabular}
	\end{center}
	\label{table:statan1}
\end{table}

\textbf{Multivariable analysis of the GRIB data.} In addition to the above metrics, we also perform multivariable analysis to measure the latency of each streaming window. The same four statistical analyses were performed but with a varying number of variables. These analyses were serialized, thus increasing the number of transformations and actions for each additional variable. There was 50GB data initially stored in HDFS which required longer execution time as each worker had to process more data. Each streaming window was once again fed with four different datasets of size 180MB, 500MB, 1GB and 2GB.

\begin{figure}[!ht]
	\begin{center}
		\includegraphics[scale=0.85]{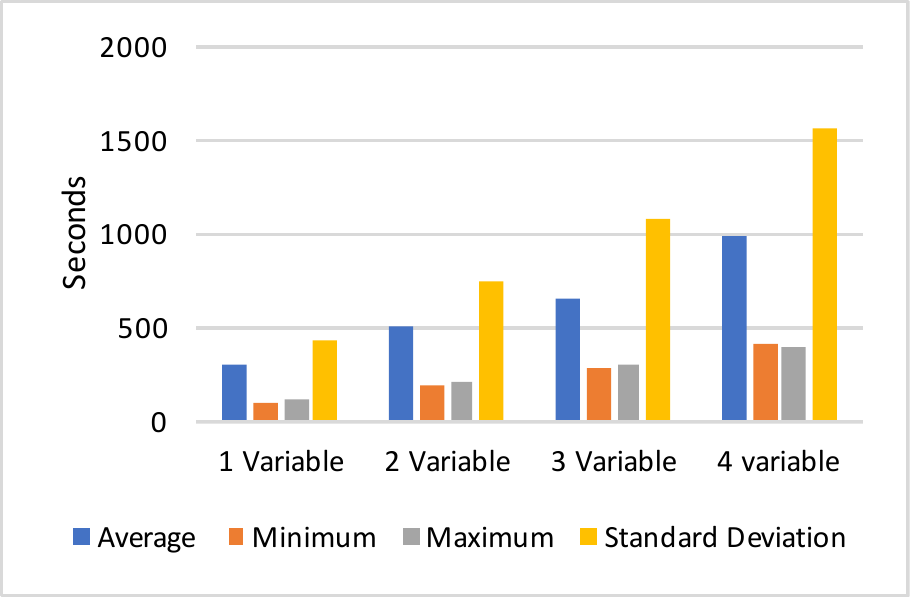}
		\caption{Statistical Analysis of the initial set of data.}
		\label{fig:statanalysis}
	\end{center}
\end{figure}

%%\noindent \includegraphics*[width=3.65in, height=2.40in, keepaspectratio=false]{image7}

%\noindent \textit{Figure 7:Statistical Analysis of the initial set of data}

%\noindent 

%When the first collect was called, it took a while to generate results. This behavior was expected as it needed to perform operations on a massive set of data. 
Figure \ref{fig:statanalysis} shows our results on the initial set of data. As expected, the execution time increases with the complexity of operation.  The \textit{standard deviation} operation took the most amount of time because the algorithm had multiple transformations to perform. 

\iffalse
Table \ref{table:std_trans} summarizes the algorithm for calculating one standard deviation for an entire set of data.

%%\noindent \includegraphics*[width=2.09in, height=0.69in, keepaspectratio=false]{image8}

%\noindent 
\begin{table}
	\caption{Standard Deviation and Transformations}
	\begin{tabular}{|p{0.8in}|p{1.6in}|} \hline 
		Transformation & Details \\ \hline 
		1 & Calculate mean for latitudes \\ \hline 
		2 & Calculate mean for longitudes \\ \hline 
		3 & Calculate the (sum of variance) * (sum of variance) \\ \hline 
		4 & Calculate the sum of variance \\ \hline 
		5 & Calculate the square root of the sum of variance / N \\ \hline 
	\end{tabular}
	\label{table:std_trans}
\end{table}
%\textit{Table 2: Standard Deviation and Transformations}
\fi

%Further, Table \ref{table:dag} shows the difference in the DAG lengths for mean, min/max, and standard deviation. It shows that the length of the DAG is directly related to the latency of the streaming job. In other words, more map and filter functions were run on the dataset, so each parallel task needed to accomplish more for operations with higher complexity.

Further, as for DAG lengths, \textit{standard deviation} has a larger DAG than those of \textit{max} and \textit{mean}. The length of a DAG is directly related to the latency of the corresponding streaming job. In other words, more map and filter functions are run on the dataset for operations with higher complexity.

%\noindent 
\iffalse
\begin{table}
	\caption{DAG visualization for each operation}
	\begin{tabular}{|p{0.7in}|p{0.7in}|p{1.0in}|} \hline 
		Mean & Min/Max & Standard Deviation \\ \hline 
		\includegraphics*[width=0.87in, height=1.60in, keepaspectratio=false]{Figs/T3_1.pdf} & \includegraphics*[width=0.90in, height=1.12in, keepaspectratio=false]{Figs/T3_2.pdf} & \includegraphics*[width=1.37in, height=3.98in, keepaspectratio=false]{Figs/T3_3.pdf} \\ \hline 
	\end{tabular}
	\label{table:dag}
\end{table}
%\textit{Table 3: DAG visualization for each operation}
\fi
%\noindent 

%\noindent The following chart shows the average time taken for different statistical analysis using a different number of variables.

\begin{figure}[!ht]
	\begin{center}
		\includegraphics[scale=0.85]{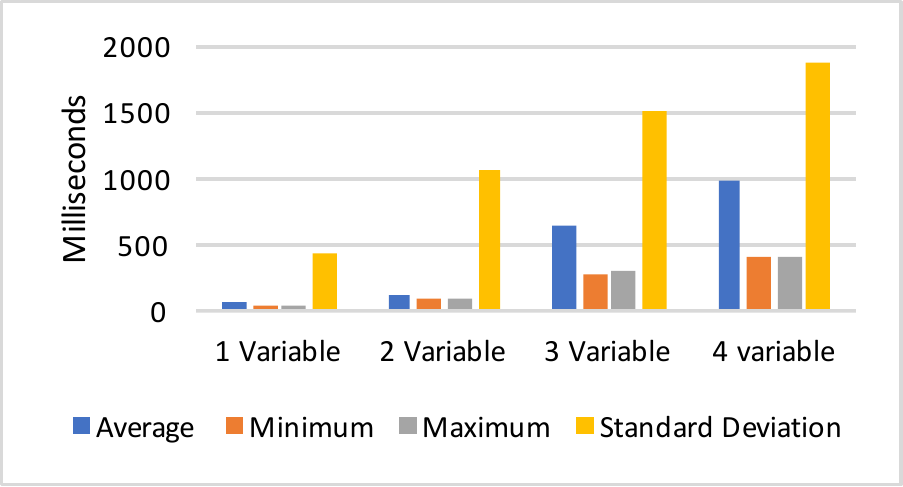}
		\caption{Statistical Analysis of batches of stream.}
		\label{fig:statanalysis2}
	\end{center}
\end{figure}

%%\noindent \includegraphics*[width=3.62in, height=1.96in, keepaspectratio=false]{image12}

%\noindent \textit{Figure 8:Statistical Analysis of batches of stream}

The size of the dataset for each 6-hour period was roughly $1$ GB in size and latency for streaming $1$ GB data was significantly smaller than the initial data. For input sources that generate discrete data at a regular interval, the streaming job is more suitable than a batch processing job because of the lack of overhead in running an iterative job \cite{RN15}. Figure \ref{fig:statanalysis2} shows the results for batches of streams, which achieves better runtime performance than the initial set of data.

\subsection{ Number of Executor Nodes }

\begin{figure}[!ht]
	\begin{center}
		\includegraphics[scale=0.85]{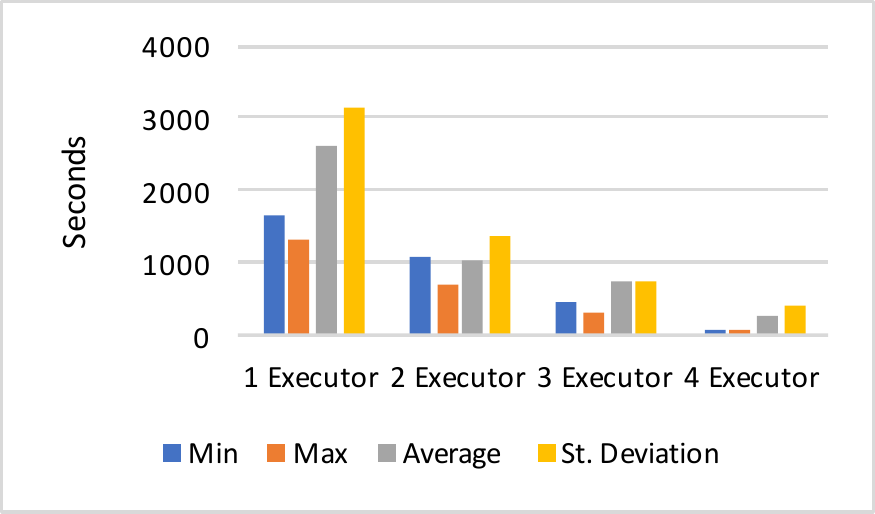}
		\caption{ Statistical analysis on initial Transformation vs. \# of executors.}
		\label{fig:statanalysis3}
	\end{center}
\end{figure}

We ran streaming jobs with a different number of worker nodes to record the change in latency. Data was streamed from HDFS and YARN was used as a dashboard to visualize states of different worker nodes. Since Apache Spark utilizes the in-memory datasets \cite{RN4}, the multi-node setup outperformed the single node-setup as it could use more resources from each worker as shown in Figure \ref{fig:statanalysis3}. It is clear from the figure that there is linear scalability in latency for a streaming job. This result shows that the efficiency of a streaming job is directly proportional to the number of workers. 
%Fig. \ref{fig:statanalysis4} shows scaling with increasng number of executors for stream of data (unlike the initial set). Similar to our previous observation, the first streaming window had the initial dataset of size 10GB, so it took longer to process that initial set.

%Similar to our previous observation, the first streaming window had the initial dataset of size 10GB, so it took longer to process that initial set. It is clear from the chart below that there is linear scalability regarding latency for a streaming job. Based on the results from this statistical analysis, one can concur that the efficiency of a streaming job would be directly proportional to the number of workers.

\subsection{ Scalability}

\begin{figure}[!ht]
	\begin{center}
		\includegraphics[scale=0.95]{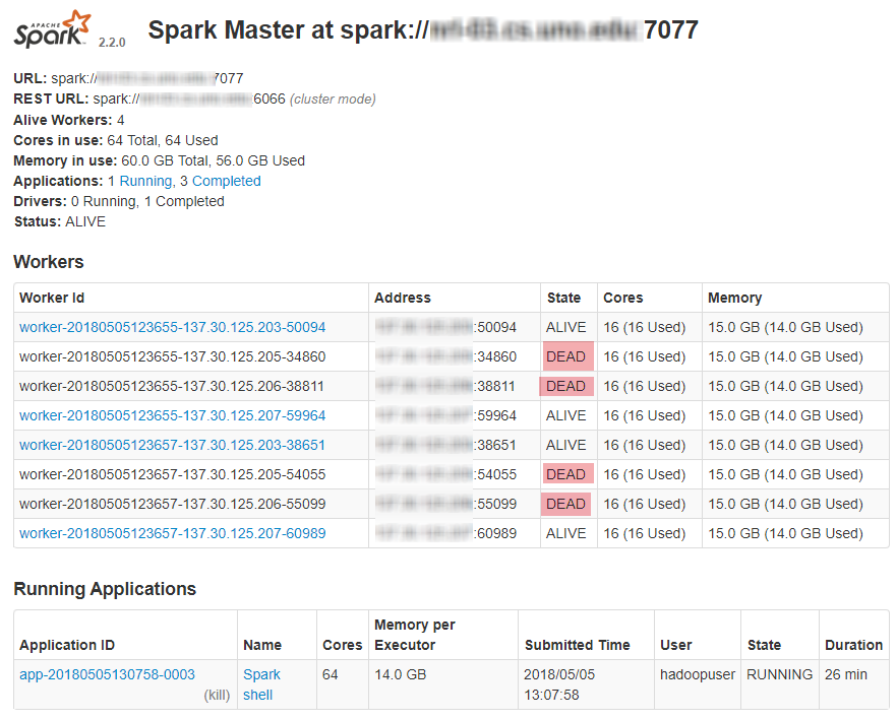}
		\caption{ Status of dead workers on YARN dashboard.}
		\vspace{-0.2in}
		\label{fig:status_worker}
	\end{center}
\end{figure}

\begin{figure}[!ht]
	\begin{center}
		\includegraphics[scale=0.95]{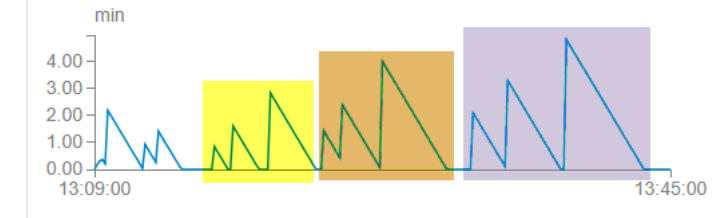}
		\caption{ Scheduling delay for different datasets with the varying number of executors. Yellow represents six executors, brown five executors, and purple three executors.}
		\label{fig:sched_delay}
	\end{center}
\end{figure}

\begin{figure}[!ht]
	\begin{center}
		\includegraphics[scale=1.05]{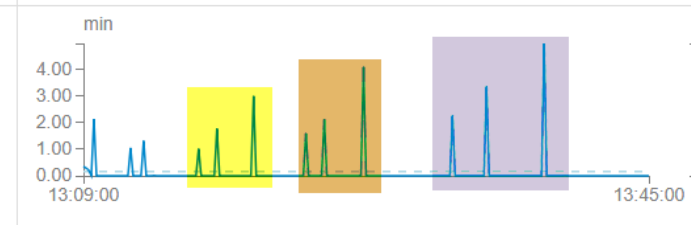}
		\caption{ Processing time for different datasets with the varying number of executors. Yellow represents six executors, brown five executors, and purple three executors.}
		\label{fig:proc_time}
	\end{center}
\end{figure}

%\noindent Another important aspect of any big data processing engine is scalability concerning both data storage and computational power [16]. 
As data grows and higher processing speed is desired, new nodes should be easily added to the cluster. During our experiment, nodes were killed and started in the cluster with a fair speed and easiness. We wrote bash scripts to control the state of a node and used YARN dashboard to verify the state. Figure \ref{fig:status_worker} shows the state of the cluster after multiple nodes were killed. Further, Figure \ref{fig:sched_delay} and \ref{fig:proc_time} showcase how different metrics of a streaming job can be visualized using Apache Spark's dashboard. Figure \ref{fig:sched_delay} plots the scheduling delay and Figure \ref{fig:proc_time} plots the processing time for batches ran with the different number of executor nodes. Yellow represents six executors, brown five executors, and purple three executors. The sizes of datasets in different batches were 180MB, 500MB, and 1GB. The scheduling delay and processing time are both directly proportional to the size of the data and inversely proportional to the number of worker instances.

%%\noindent \includegraphics*[width=3.58in, height=2.88in, keepaspectratio=false]{image16}

%\noindent \textit{Figure }12\textit{:Showing the status of dead workers on YARN dashboard}

%\noindent 

%\noindent 

%%\noindent \includegraphics*[width=3.86in, height=0.92in, keepaspectratio=false]{image17}

%\noindent \textit{Figure 13:Scheduling delay for different dataset with the varying number of executors}

%\noindent 

%%\noindent \includegraphics*[width=3.86in, height=1.00in, keepaspectratio=false]{image18}

%\noindent \textit{Figure 14:Processing Time for different dataset with the varying number of executors}

\subsection{ Fault Tolerance }

\begin{figure}[!ht]
	\begin{center}
		\includegraphics[scale=0.95]{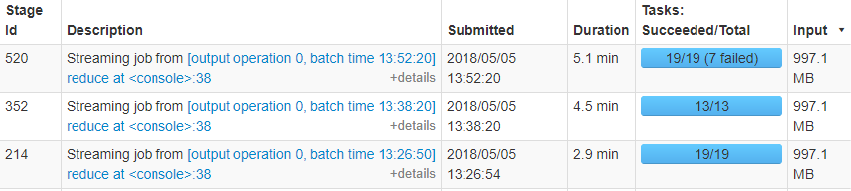}
		\caption{ Difference in processing time for node failures. The first row demonstrates the execution time with node failures.}
		\label{fig:diff_time}
	\end{center}
\end{figure}

\noindent Spark can reconstruct the RDDs using lineage information stored in the RDD objects when a node falls apart \cite{RN4}. Since the data is already replicated across nodes in HDFS, lost partitions can be reconstructed in parallel across multiple nodes without much overhead. If the node running receiver fails, then another node is spun up with the receiver which can continue to read from HDFS. If the receiver was using Kafka or Flume as a source instead of HDFS, then a small amount of data may be lost which hasn't been replicated to other nodes in the cluster \cite{RN15}. We measure performance of a system running streaming job with various node failures to access the fault tolerance capability of the Apache Spark streaming. Spark's dashboard interface was used to visualize the difference in latency for different batches running with and without node failures. Figure \ref{fig:diff_time} shows that if some nodes fail while running a batch, it will take longer to account for the lost nodes and reschedule those jobs in different node/s. For instance, stage ID $520$ lost a node with two workers, and the driver had to reschedule seven tasks running on that node somewhere else. As a result, the latency increased from $2.9$ to $5.1$ minutes.
\iffalse
The dashboard also showed the details about the failure which could aid in debugging the issue during unexpected failures.

\begin{figure}[!ht]
	\begin{center}
		\includegraphics[scale=0.85]{Figs/Fig15_exp.pdf}
		\caption{ \small Explanation for a failing nodes}
		\label{fig:exp_fail}
	\end{center}
\end{figure}

%%\noindent \includegraphics*[width=7.38in, height=0.69in, keepaspectratio=false]{image19}

%\noindent \textit{Figure 15:Explanation for a failing node}

%\noindent 
\fi

%%\noindent \includegraphics*[width=3.41in, height=0.77in, keepaspectratio=false]{image20}

%\noindent \textit{Figure 16:Difference in processing time for node failures}

\subsection{ VISUAL APPLICATION}
We develop a web interface to demonstrates a sample usage of our application. The web page uses Google Maps and its developer API to visualize the results generated by our application. The web application is written in .NET MVC framework. The server-side code grabs the latest result from the cluster by using the WinSCP library (this was used to avoid installing FTP on the master in the cluster), then converts the results into text format using ncl\_dump. A text dump of the resulting NetCDF file was processed and sent to the view. A JavaScript function regularly polls for the result, and once the Spark application generates the result, it is visualized on the web.

\iffalse
\noindent The following figure summarizes the workflow of the web application:

\begin{figure}[!ht]
	\begin{center}
		\includegraphics[scale=0.75]{Figs/Fig17_summary.pdf}
		\caption{Summary of the web application}
		\vspace{-0.2in}
		\label{fig:summ_web}
	\end{center}
\end{figure}
\fi

%%\noindent \includegraphics*[width=3.60in, height=1.50in, keepaspectratio=false]{image21}

%\noindent \textit{Figure 17:Summary of the web application}

%\noindent 

\begin{figure}[!ht]
	\begin{center}
		\includegraphics[scale=0.85]{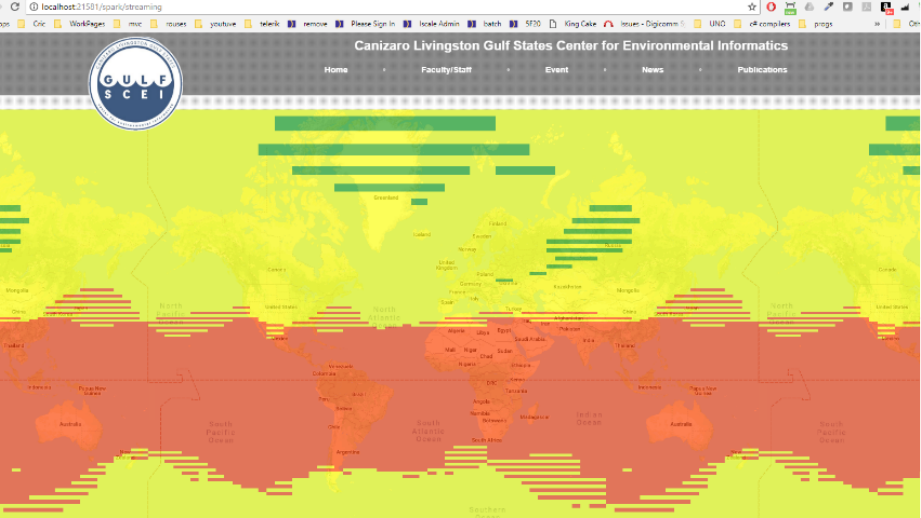}
		\caption{ Screenshot of color-coded representation of the result. This UI visualizes a single variable result file by color coding the latitude and
			longitude over google map based on the value of the variable for that coordinate.}
		\label{fig:screen_result}
	\end{center}
\end{figure}

\section{ Additional Observations and Findings}

\textbf{Spark Streaming vs. Hadoop's batch processing vs. Storm Trident.} An iterative job like the one used in this experiment can be expressed as multiple Map and Reduce operations in Hadoop. However, different MapReduce jobs cannot share data. So for iterative analysis, the same dataset must be read from HDFS multiple times, and results would need to be written to HDFS many times as well \cite{RN17}. These iterations create much overhead because of the I/O operations and other unwanted computations \cite{RN18}. Spark tackles these issues by storing intermediate results in memory. Spark Streaming uses D-Streams or discretized streams of RDDs which provides consistent, ``exactly-once" processing across the cluster \cite{RN19} and thus significantly increases the performance for iterative analysis. Apache Storm can process unbounded streams of data in real time, and it can be used alongside Hadoop, but it only guarantees ``at-least-once" processing \cite{RN20}. Trident bolsters Storm by providing micro-batching and other abstractions that would ensure ``exactly-once" processing \cite{RN21}. It would take three different libraries to work seamlessly to accomplish what Spark Streaming can accomplish by itself. Time and effort required to setup and maintain Storm Trident application along with Hadoop can hamper the production and deployment. In contrast, Spark's Streaming library is directly written over its core and maintained by the same group who maintain the core's code base. Thus, Spark streaming outshines both Hadoop and Storm Trident combination for streaming scientific data.

\iffalse
\subsection{Limitations of Spark}

\noindent The limitation of using Spark over Hadoop boils down to memory. When a dataset is large enough not to allow any more RDDs to be stored in memory, Sparks starts to replace RDDs, and such frequent replacement degrades the latency and thus makes Hadoop more suitable in such situations \cite{RN22}. When running spark streaming job with two nodes and four workers with each node getting maximum of 2GB memory, Spark crashed and threw JVM heap exceptions. So, when memory is not abundant for large datasets, batch processing in Hadoop is a better choice. However, for this research, we needed a framework that would seamlessly stream datasets that were only relatively large, and Spark Streaming was able to handle it efficiently.
\fi

\textbf{Limitations of Spark.} When a dataset is large enough not to allow any more RDDs to be stored in memory, Sparks starts to replace RDDs, and such frequent replacement degrades the latency \cite{RN22}. However, for this work, we needed a framework that would seamlessly stream a relatively large datasets, and Spark Streaming was able to handle it efficiently.

\section{ CONCLUSIONS}

We use SciSpark successfully with Apache Spark to stream GRIB1 data in a streaming application. The bulk of the logic in this application lies in the ability to convert the statistical analysis into transformations and actions that would run upon the DStream of RDDs of type SciTensor. Datasets ranging from 180MB to 50GB were used in the application without running into any memory issues. Various properties of a streaming application like operation complexity, scalability and fault tolerance were assessed, and results were summarized using simple mathematical operations like mean, min/max and standard deviation. Based on these results and other properties of apache Spark Streaming, we are confident that Spark Streaming is a better solution to stream the scientific data over Hadoop or Storm Trident. 

\iffalse
\noindent 

\noindent ACKNOWLEDGEMENT

\noindent I would like to present my gratitude towards my supervisor Dr. Mahdi Abdelguerfui for his continuous support and guidance at every level throughout the process of this Thesis work. I would like to thank Dr. Elias Ioup for his valuable advice and knowledge about the direction of this project, and Dr. Shaikh M Arifuzzaman and Dr. Tamjidul Hoque for reviewing the manuscript and suggesting changes. I would like to appreciate Dr. Vassil Roussev and Dr. Shengru Tu for their productive coursework in different classes relating to distributed systems.

\noindent I would also like to present my appreciation towards all my coworkers and lab partners at Cannizaro Livingston Gulf States Center for Environmental Informatics (GulfSCEI).

\noindent As an effort to further grow this research, I would like to take these findings and use them to solve real-world problems.
\fi

\section*{ACKNOWLEDGMENT}
\small
Part of this work was supported by ONR contracts N00173-16-2-C902 and N00173-14-2-C901, and Louisiana Board of Regents RCS Grant LEQSF(2017-20)-RD-A-25.

\bibliographystyle{scsproc}
% AUTHOR: Include your bib file here
\bibliography{spark.bib,demobib}

\section*{Author Biographies}

\textbf{\uppercase{Janak Dahal}} got his Masters degree in Computer Science from the University of New Orleans in 2018. His research interests lie in large-scale data analysis and systems development. His email address is \email{jdahal@uno.edu}.

\textbf{\uppercase{Elias Ioup}} is with US Naval Research Laboratory, Stennis Space Center, MS, USA. His research interests are in Spatio-temporal Database Systems, High-performance Geographical Information Systems, and Large-scale Systems. His email address is \email{elias.ioup@nrlssc.navy.mil}.

\textbf{\uppercase{Shaikh Arifuzzaman}} is an Assistant Professor in Computer Science Department at the University of New Orleans. He holds a Ph.D. in Computer Science from Virginia Tech. His research interests include Big Data, High Performance Computing, Parallel Algorithms, and Graph Mining. His email address is \email{smarifuz@uno.edu}.

\textbf{\uppercase{Mahdi Abdelguerfi}} is the Chairman of Computer Science Department at the University of New Orleans. He is also the director of Canizaro Livingston Gulf States Center for Environmental Informatics (GulfSCEI). His research interests are in Spatio-temporal Database Systems, High-performance Geographical Information Systems, Synthetic Environments Reconstruction and Parallel Database Systems. His email address is \email{mahdi@cs.uno.edu}.

\end{document}